\documentclass[11pt,a4paper]{article}

\usepackage[utf8]{inputenc}
\usepackage[T1]{fontenc}
\usepackage[margin=2.5cm]{geometry}
\usepackage{authblk}
\usepackage{tikz}
\usepackage{amsmath,amssymb}
\usepackage{physics}
\usepackage{graphicx}
\usepackage{subcaption}
\usepackage{booktabs}
\usepackage{multirow}
\usepackage{xcolor}
\usepackage[colorlinks=true,allcolors=blue]{hyperref}
\usepackage[backend=biber,style=ieee]{biblatex}
\title{Soft Mixture of Quantum Experts\\for Hybrid Quantum--Classical Learning}

\author[1,2]{Valeria Repetto\thanks{Corresponding author: valeria.repetto@unito.it}}
\author[1]{Romina D'Aurizio}
\author[2]{Piero Fariselli}

\affil[1]{National Research Council of Italy (CNR), Institute of Informatics and Telematics (IIT), via Giuseppe Moruzzi 1, Pisa, Italy}
\affil[2]{Computational Biomedicine Unit, Department of Medical Sciences, University of Turin, via Santena 19, Torino, Italy}

\date{}

\begin{document}

\maketitle

\begin{abstract}

Quantum Machine Learning (QML) has attracted increasing interest as a possible route to construct feature representations that are difficult to reproduce with classical models. However, near-term QML architectures based on trainable parametrized quantum circuits remain affected by limited hardware resources, noise, and severe optimization difficulties such as barren plateaus. In this work we propose Soft-MoQE, a Soft Mixture of Quantum Experts architecture that replaces variational quantum training with the input-dependent selection and combination of fixed quantum feature maps. The model is composed of a heterogeneous pool of non-trainable quantum circuits, each defining a distinct embedding, and a classical gating network that learns how to weight these embeddings for each data point. We benchmark Soft-MoQE on three  PCA-reduced tasks: MNIST parity, MNIST 10-class, and FashionMNIST. Soft-MoQE outperforms the single best expert, the uniform average of all experts and a linear baseline, achieving task-dependent performance comparable to that of a trainable quantum neural network, without optimizing any quantum parameter and with a quantum training cost that is lower by orders of magnitude. The analysis of the class-conditioned expert usage shows that the learned routing combines a few experts for each input, with a selection that depends on the class. These results support the learned routing among fixed quantum circuits as a trainability-aware design principle for hybrid QML.

\end{abstract}

\section{Introduction}
\label{introduction}

Quantum Machine Learning (QML) has emerged as a promising paradigm for developing data-driven algorithms through quantum computation \cite{Biamonte2016QuantumML, Dunjko2016QuantumenhancedML}. By embedding classical data into high-dimensional Hilbert spaces and exploiting quantum superposition and entanglement, quantum models may induce feature representations that are difficult to simulate classically~\cite{Schuld2018QuantumML}. However, in the current Noisy Intermediate-Scale Quantum (NISQ) era, quantum hardware remains limited by noise, restricted qubit counts, and the absence of full fault tolerance~\cite{PerdomoOrtiz2017OpportunitiesAC}.  In this context, variational models, in which parametrized quantum circuits (PQCs) are trained by classical optimizers~\cite{mcclean2016theory, benedetti2019parameterized}, have become the most widely adopted approach \cite{cerezo_variational_2021}.  However, training remains a major obstacle since a broad class of PQCs suffer from barren plateaus, in which the variance of the gradient vanishes exponentially with the number of qubits \cite{mcclean_barren_2018, cerezo_cost_2021}. In such regimes, optimization landscapes become dominated by flat regions, rendering gradient-based training increasingly impractical as circuit size grows. Moreover, on quantum hardware, the estimation of the gradient requires a number of circuit executions that grows with the number of trainable parameters ~\cite{abbas2023quantum}.
These limitations have motivated a shift beyond variational models, in which the optimization is moved away from the quantum parameters \cite{huang_post-variational_2024}. In post-variational approaches, a set of fixed quantum circuits is measured and only the classical combination of their outputs is trained, which avoids the optimization of quantum parameters while retaining quantum representations. Combining multiple quantum models has also been explored through quantum ensemble methods, in which the predictions of several quantum classifiers are aggregated with weights that do not depend on the input \cite{schuld_quantum_2018, incudini2023resource, chen2025quantum}. Since the most suitable encoding depends on the data and on the task \cite{Schuld2021QuantumML, Kbler2021TheIB} , a natural question is how to select and weight fixed quantum representations according to each input.

The Mixture of Experts (MoE) paradigm \cite{Jacobs_MOE} offers a natural framework to address this question. In a MoE, several specialized submodels are combined through a gating network that weights them on a per-input basis, so that different inputs can rely on different experts. Recent works have extended the MoE paradigm to fully quantum settings, where classical experts are replaced by parametrized quantum neural networks and the gating mechanism itself is implemented through quantum circuits \cite{tognini_solving_2025, nguyen_qmoe_2025}. While these approaches report promising performance, they rely on fully trainable quantum components and have so far been evaluated on binary classification tasks. Moreover, the simultaneous optimization of multiple quantum experts together with a quantum gating network entails substantial computational overhead and exacerbates trainability issues inherent to PQCs.

In this work we introduce the Soft Mixture of Quantum Experts (Soft-MoQE), in which the experts are fixed quantum feature maps and the routing is learned by a classical gating network. Each expert implements a different encoding, from single-qubit rotations to entangling Instantaneous Quantum Polynomial(IQP) circuits \cite{Havlek2018SupervisedLW}, and the gating network combines their outputs through a temperature-scaled softmax, with weights that adapt to each input. The model thus adapts the quantum representation to the data while keeping all trainable parameters classical, so that no gradient with respect to quantum parameters is required and barren plateaus do not arise by construction. We evaluate Soft-MoQE on MNIST \cite{lecun2010mnist} and FashionMNIST \cite{xiao2017fashionmnist} after Principal Component Analysis (PCA)  reduction, on a parity task and on 10-class classification. The learned routing outperforms both the best individual expert and the uniform average of all experts, and the gating weights reveal a selective, class-dependent use of the encodings. Compared with a variational Quantum Neural Network(QNN) acting on the same input, Soft-MoQE reaches comparable, task-dependent performance without training any quantum parameter and with a quantum cost of training lower by orders of magnitude.
Overall, our results indicate that modular, fixed quantum representations combined through classical mixture mechanisms constitute a viable and trainability-aware design principle for hybrid learning in the NISQ era. 

\section{Mixture of Quantum Experts}

\subsection{Model architecture}

The Mixture of Experts (MoE) framework is an ensemble learning approach in which a model is composed of multiple specialized submodels, referred to as experts, each of which is responsible for learning different aspects of the data or solving different sub-tasks. Rather than combining expert outputs uniformly, MoE employs a gating mechanism that dynamically determines which experts should contribute, and how strongly they should contribute, to the final prediction for a given input \cite{cai_survey_2025}. 
Formally, a MoE consists of a set of $n$ experts $\{f_1, \dots, f_n\}$, where each expert $f_i : \mathcal{X} \to \mathcal{Y}$ receives the same input $x \in \mathcal{X}$. A gating network $G$ maps the input $x$ to a vector of non-negative weights that sum to one:
\begin{equation}
G(x) = \left(w_1(x), \dots, w_n(x)\right), \quad w_i(x) \ge 0, \quad \sum_{i=1}^{n} w_i(x) = 1.
\label{eq:gate}
\end{equation}
The output of the MoE is the weighted combination of the expert outputs,
\begin{equation}
f(x) = \sum_{i=1}^{n} w_i(x)\, f_i(x).
\label{eq:moe}
\end{equation}

\begin{figure*}[!t]
    \centering
    \includegraphics[width=1\linewidth]{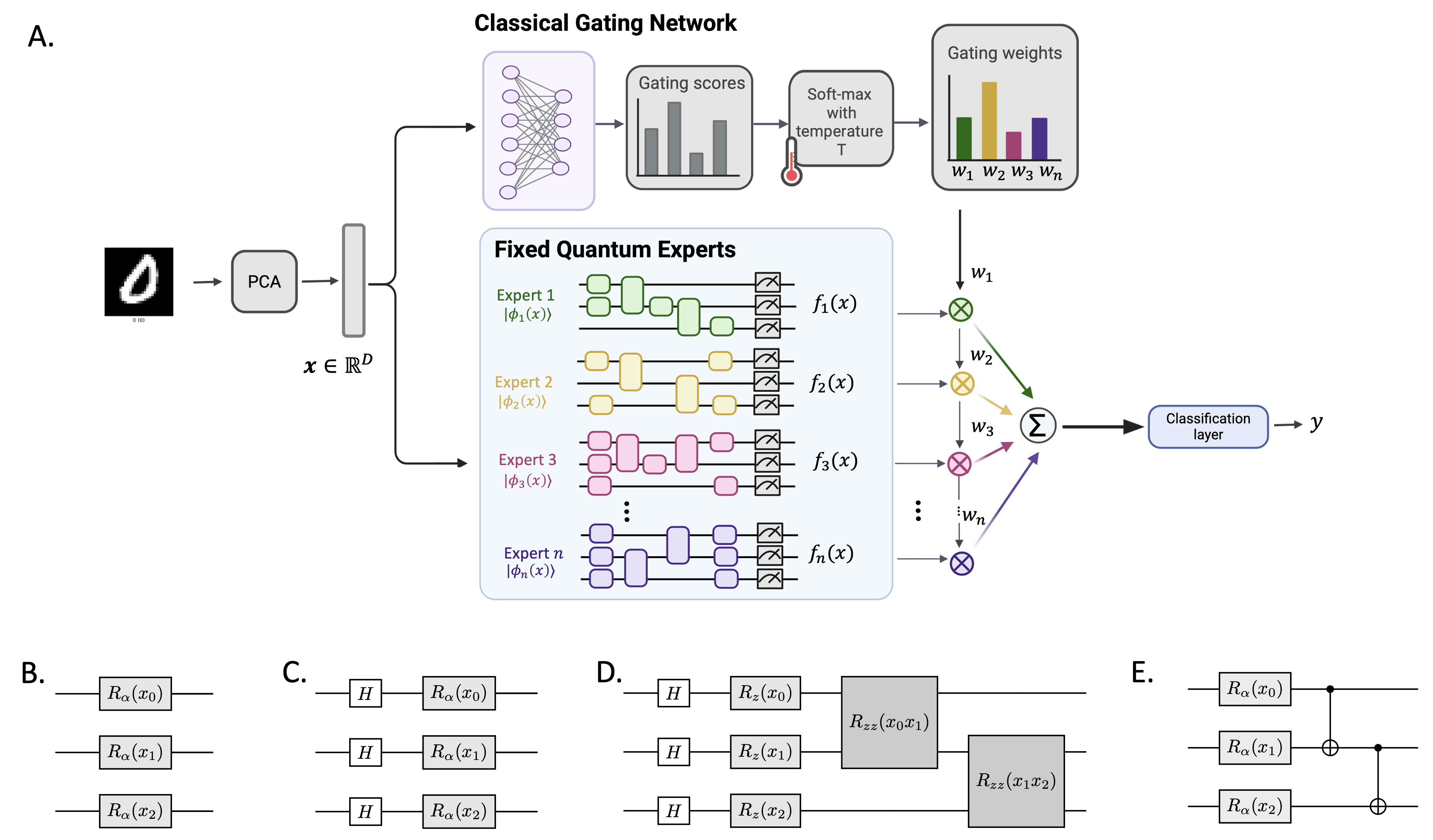}
    \caption{Soft-MoQE architecture and quantum experts. (A) The input is reduced by PCA to $x \in \mathbb{R}^D$ and passed both to the classical gating network and to the $n$ fixed quantum experts. The gating network assigns a score to each expert, and a softmax with temperature $T$ converts the scores into the weights $w_i$. Each expert prepares the state $|\phi_i(x)\rangle$, whose measurement gives the vector of expectation values $f_i(x)$. The weighted sum $\sum_i w_i f_i(x)$ is passed to the classification layer, which returns the prediction $y$. (B--E) Encoding circuits of the four expert families, shown on three qubits for a single repetition. (B) Rotational embedding, (C) Hadamard embedding, (D) IQP embedding with linear entanglement, whose full variant couples all pairs of qubits, and (E) entangled rotational embedding with a linear ladder of CNOT gates. Here $\alpha \in \{X, Y\}$ denotes the rotation axis in (B), (C) and (E), while in (D) $\alpha = Z$. In the experiments each circuit acts on $D = 10$ qubits and is repeated $r$ times, with the input re-uploaded at each repetition.}
    \label{fig:graphabs_circ}
\end{figure*}

The outputs of the experts are typically aggregated with a weighted sum or by combining the top experts selected by $G$.
Building on this classical paradigm, we propose Soft-MoQE, in which the experts are implemented as quantum circuits without trainable parameters. Each quantum expert defines a quantum feature map $\phi_i$ that embeds the input $x$ into a high-dimensional Hilbert space. Different experts employ distinct quantum embeddings, allowing the model to exploit diverse quantum representations of the input data. A detailed description of the quantum experts and their feature maps is provided in Section~\ref{sec:experts}. In this setting, the gating network learns to select and combine quantum embeddings on an input-dependent basis, and the resulting weighted representation is fed to a final output layer. The model architecture is shown in Figure~\ref{fig:graphabs_circ} A.
The gating function $G(x)$ is implemented as a multilayer perceptron (MLP) with a single hidden layer of size $h$ and ReLU activation, which produces a set of unnormalized scores $\{z_1(x), \dots, z_n(x)\}$. These scores are converted into expert weights through a temperature-scaled softmax function,
\begin{equation}
w_i(x) = \frac{\exp\left(z_i(x)/T\right)}{\sum_{j=1}^{n} \exp\left(z_j(x)/T\right)},
\label{eq:softmax}
\end{equation}
where $T > 0$ denotes the temperature parameter. High temperatures ($T > 1$) should favour a more uniform allocation of weights across experts, whereas low temperatures ($T < 1$) lead to a more peaked distribution, so that the gating network focuses on a smaller subset of experts. From a statistical mechanics perspective, this weighting scheme corresponds to a Boltzmann distribution over experts, where the scores $z_i$ play the role of negative energies. Since the scores are produced by a trainable network, $T$ acts mainly on the initial sharpness of the distribution and on the training dynamics, rather than imposing a fixed level of selectivity.
The weighted quantum embeddings are then combined and passed to a final classical prediction layer $F$,
\begin{equation}
y(x) = F\left( \sum_{i=1}^{n} w_i(x) f_i(x) \right).
\label{eq:output}
\end{equation}
In this way, the gating network learns to route different inputs to the most relevant quantum embeddings while preserving the end-to-end differentiability of the model. In all experiments $F$ is a single linear layer followed by a softmax. The only trainable components are therefore classical, namely the gating network and the output layer. The quantum circuits receive the reduced input directly, without any trainable classical preprocessing, so that the representation passed to the output layer is built only from the expert outputs.

\subsection{The quantum experts}
\label{sec:experts}

The quantum experts in the proposed Soft-MoQE framework are designed as fixed quantum feature maps instead of trainable parametrized quantum circuits. This choice allows the model to exploit the expressive power of quantum embeddings while keeping the training of the overall architecture fully classical. Each expert corresponds to a distinct quantum mapping $\phi_i$ that projects input data points into quantum states, inducing a different representation in the Hilbert space. A classical data point $x \in \mathbb{R}^D$ is mapped into a quantum state $|\phi_i(x)\rangle$ by applying a unitary transformation $U_{\phi_i}(x)$ acting on $D$ qubits initialized in $|0\rangle^{\otimes D}$,
\begin{equation}
\phi_i : x \longrightarrow |\phi_i(x)\rangle = U_{\phi_i}(x) |0\rangle^{\otimes D}.
\label{eq:featuremap}
\end{equation}
Information is extracted from the resulting quantum state by measuring expectation values of Pauli observables, which constitute the output of each quantum expert,
\begin{equation}
f_i(x) = \left(\langle Z_1 \rangle, \langle Z_2 \rangle, \dots, \langle Z_D \rangle\right).
\label{eq:expert_output}
\end{equation}
These outputs are subsequently combined by the gating mechanism and passed to the final classical prediction layer (see Eq.~\ref{eq:output}).

The unitary transformations $U_{\phi_i}$ can be constructed in multiple ways, leading to quantum feature maps of varying expressive complexity \cite{schuld2021effect,Kbler2021TheIB, larose2020robust}. To provide Soft-MoQE with heterogeneous levels of representation, we employ four families of encodings, shown in Fig.~\ref{fig:graphabs_circ}. The first two use only single-qubit gates. Rotational embeddings consist of layers of rotations $R_\alpha(x_j)$ about the axis $\alpha \in \{X, Y\}$, one per qubit (Fig.~\ref{fig:graphabs_circ} B), while Hadamard embeddings combine each rotation layer with a layer of Hadamard gates (Fig.~\ref{fig:graphabs_circ}C). The other two families introduce entanglement. IQP-style encodings \cite{Havlek2018SupervisedLW} consist of a layer of Hadamard gates followed by a diagonal unitary $U_Z(x)$, composed of single-qubit rotations $R_Z(x_j)$ and two-qubit rotations $R_{ZZ}(x_j x_k)$, the latter acting either between neighbouring qubits (linear) or between all pairs of qubits (full). Finally, entangled rotational embeddings alternate a layer of rotations $R_\alpha(x_j)$, $\alpha \in \{X, Y\}$, with a linear ladder of CNOT gates acting on neighbouring qubits (Fig.~\ref{fig:graphabs_circ} E).

In addition to varying the type of quantum encoding, we further diversify the experts by employing different numbers of repetitions $r \in \{2, 3, 4\}$ of the corresponding circuit, with the input re-uploaded at each repetition. Each quantum expert is thus characterized by its encoding structure, its entanglement pattern and its repetition depth. 
Some combinations of encoding, rotation axis and depth also produce non-informative experts. 

We therefore computed all candidate experts and removed those whose outputs were constant over the training set or identical to the outputs of another expert. The resulting library contains the 19 experts listed in Table~\ref{tab:experts}.

\begin{table}[t]
\centering
\small
\setlength{\tabcolsep}{4pt}
\caption{Library of quantum experts used in Soft-MoQE after removing constant and duplicate experts. $r$ denotes the number of repetitions of the encoding circuit.}
\label{tab:experts}
\begin{tabular}{@{}lllcc@{}}
\toprule
Family & Entanglement & Axis & $r$ & Experts \\
\midrule
Rotational & none & $X$ & 2, 3, 4 & 3 \\
Hadamard & none & $X$ & 2, 3, 4 & 3 \\
 &  & $Y$ & 3 & 1 \\
IQP & linear & & 2, 3, 4 & 3 \\
 & full & & 2, 3, 4 & 3 \\
Entangled rotational & CNOT ladder & $X$, $Y$ & 2, 3, 4 & 6 \\
\midrule
Total & & & & 19 \\
\bottomrule
\end{tabular}
\end{table}

\section{Experimental setup}
\label{sec:setup}

\subsection{Datasets and preprocessing}
\label{sec:data}

The proposed model has been benchmarked on the MNIST dataset \cite{lecun2010mnist}, which consists of grayscale images of handwritten digits (0 to 9) of size $28 \times 28$ pixels, and on the FashionMNIST dataset \cite{xiao2017fashionmnist}, which has the same image size and number of classes but contains clothing categories. Both datasets contain 60000 training images and 10000 test images. For model optimization we further divided the training set into 50000 training images and 10000 validation images. On MNIST we consider both the 10-class digit classification task and a parity task, in which digits are labelled as even or odd. To accommodate the quantum encoding constraints, the images were reduced in dimensionality using PCA, retaining the first 10 principal components, so that each expert acts on $D = 10$ qubits. PCA and the subsequent  rescaling of each component to the interval $[0, \pi]$ were fitted on the training data only and then applied to the test set, clipping the values that fall outside this interval. The expert encodings of the whole dataset were precomputed with PennyLane \cite{bergholm2022pennylaneautomaticdifferentiationhybrid} by exact statevector simulation.

\subsection{Baselines}
\label{sec:baselines}

To place the performance of Soft-MoQE in context, we consider classical and quantum reference models, together with two ablations of the routing mechanism. As a classical reference, an MLP with a single hidden layer of 50 units and ReLU activation was trained on the same reduced representation, with fixed hyperparameters. Its performance serves as a practical proxy for the amount of task-relevant information retained after dimensionality reduction, rather than as a direct performance target. A linear baseline, consisting of a single-layer model identical to the final classification layer of Soft-MoQE and trained on the same reduced input, isolates the contribution of the output layer.
To distinguish the contribution of the learned gating mechanism from simple ensemble effects, we compare Soft-MoQE against two ablation baselines. In the uniform average, all experts receive the same weight $w_i = 1/n$ and their combined output is passed to the same linear layer used in Soft-MoQE. In the single best expert baseline, the linear layer is trained on the output of a single expert, selected in each sampling as the one achieving the highest validation Matthews correlation coefficient (MCC). In all samplings and tasks this was the rotational expert with $X$ rotations and $r = 2$.
As a quantum baseline we consider a variational QNN acting on the same reduced input. Each input is encoded through single-qubit $R_Y$ rotations, followed by $L$ strongly entangling layers \cite{schuld2020circuit} on 10 qubits, for a total of $30L$ trainable quantum parameters. The expectation values $\langle Z_j \rangle$ of all qubits are passed to a linear output layer identical to that of Soft-MoQE. The QNN was simulated with PennyLane through the PyTorch interface, with gradients computed by backpropagation.

\subsection{Training and model selection}
\label{sec:training}

All models were implemented in PyTorch \cite{paszke2019pytorchimperativestylehighperformance} and trained with the same protocol. Optimization was performed using the Adam optimizer~\cite{kingma2015adam} on minibatches of $1000$ samples, with early stopping based on the validation loss, a patience of $20 $ epochs and a maximum of $500$ epochs. For each model and configuration, training was repeated over the same $30$ random training/validation samplings, so that all models are evaluated on identical splits.

For Soft-MoQE we explored temperatures $T \in \{0.001, 0.1, 1, 10\}$, which control the sharpness of the expert selection distribution produced by the gating network (see Eq.~\ref{eq:softmax}), and gating hidden sizes $h \in \{20, 50\}$ with a learning rate of $10^{-3}$. For the QNN we explored $L \in \{1, 2, 4, 8\}$ and learning rates in $\{10^{-3}, 10^{-2}\}$. For each model and task, the best configuration was selected as the one with the highest mean validation MCC. Performance is reported on the test set in terms of MCC and accuracy, as mean and standard deviation over the 30 samplings.

\section{Results}
\label{sec:results}

\subsection{Model selection and stability}
\label{sec:selection}
\begin{figure*}[h!]
\centering
    \includegraphics[width=0.95\textwidth]{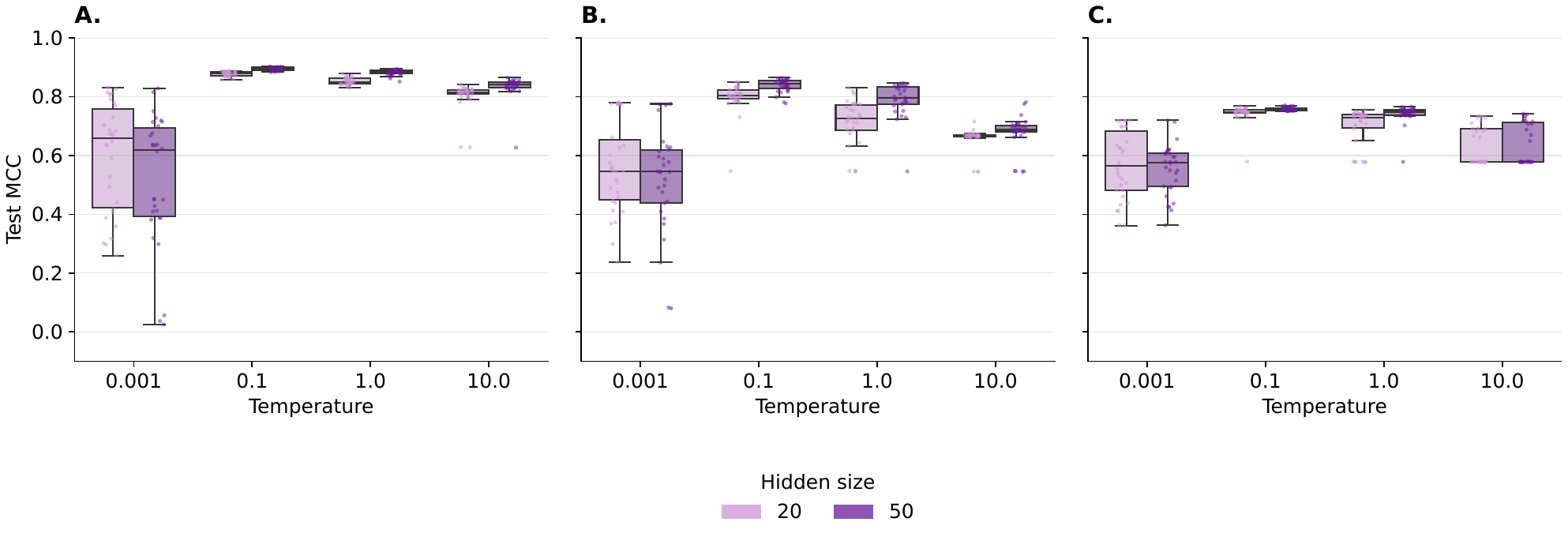}
\caption{Model selection and stability results. Box plots of test MCC across temperature, coloured by the hidden size of the gating network, for (A) parity, (B) 10-class digit classification and (C) FashionMNIST.}
\label{fig:opt}
\end{figure*}

Optimization results show that Soft-MoQE benefits from a moderately selective routing,  where the best configuration, selected according to the highest mean validation MCC, is $T = 0.1$ and $h = 50$  over all three tasks.  This configuration is adopted as the reference Soft-MoQE model for all subsequent analyses.
With this configuration, Soft-MoQE reaches an average test MCC of $0.894 \pm 0.006$ on the parity task, $0.84 \pm 0.02$ on the 10-class task and $0.758 \pm 0.006$ on FashionMNIST(Figure~\ref{fig:opt}).

The temperature also influences the stability of the model, since an overly sharp gating distribution makes the results strongly dependent on the training/validation split. Considering the gating network with $h = 50$, for $T = 0.001$ the test MCC spans from $0.080$ to $0.776$ on the parity task, from $0.025$ to $0.694$ on the 10-class task and from $0.363$  to $0.720$ on FashionMNIST, with standard deviations of $0.176$, $0.222$, and $0.087$, respectively, suggesting that an excessively sharp expert-selection distribution can negatively impact robustness and lead to inconsistent results. At such low temperatures only a very small subset of experts is selected, making the behaviour of the model highly sensitive to which specific expert is activated. Since different experts represent distinct quantum embeddings, collapsing the gating distribution onto a single expert may result in substantially different outputs depending on the selected expert. At higher temperatures this sensitivity is reduced, because the selection is more evenly distributed. The MCC test values, reported in Figure \ref{fig:opt}, show that except at $T=0.001$ the model benefits from wider classical gating networks across tasks, allowing the model to select and combine the experts more effectively.

\subsection{ Learned routing mechanism}
\label{sec:ablation study}
To distinguish  whether the contribution of the learned gating mechanism stems from simple ensemble effects or is mostly carried by a specific feature map, we compare Soft-MoQE  with the two non-adaptive baselines introduced in Section \ref{sec:baselines}, the single best expert and the uniform average of all experts.
The MCC and accuracy values reported in Table \ref{tab:aggregation_performance} show that Soft-MoQE outperforms both baselines on all the evaluated tasks. 

\begin{table*}[t]
\centering
\caption{MCC and accuracy on the MNIST parity, MNIST 10-class and FashionMNIST test sets for different aggregation strategies. Results are reported as mean $\pm$ standard deviation over 30 training/validation samplings.}
\resizebox{0.95\textwidth}{!}
{
\begin{tabular}{@{}lcccccc@{}}
\toprule
\multicolumn{1}{c}{\multirow{2}{*}{\textbf{Model}}} &
\multicolumn{2}{c}{\textbf{MNIST: Parity}} &
\multicolumn{2}{c}{\textbf{MNIST: 10-class}} &
\multicolumn{2}{c}{\textbf{FashionMNIST}} \\
\cmidrule(l){2-7}
 & MCC & Accuracy & MCC & Accuracy & MCC & Accuracy \\
\midrule

Single best expert
& $0.6374\pm 0.0007$& $0.8185\pm 0.0004$& $0.7790\pm 0.0004$& $0.8011\pm 0.0004$& $0.7199\pm 0.0007$& $0.7477\pm 0.0007$\\

Uniform average of experts
& $0.327 \pm 0.002$
& $0.662 \pm 0.001$
& $0.3752 \pm 0.0007$
& $0.4374 \pm 0.0007$
& $0.5146 \pm 0.0009$
& $0.5624 \pm 0.0008$ \\

Soft-MoQE
& $0.894 \pm 0.006$
& $0.947 \pm 0.003$
& $0.84 \pm 0.02$
& $0.85 \pm 0.02$
& $0.758 \pm 0.006$
& $0.782 \pm 0.005$ \\

\bottomrule
\end{tabular}
}
\label{tab:aggregation_performance}
\end{table*}

\begin{figure*}[h!]
\centering
    \includegraphics[width=0.95 \textwidth]{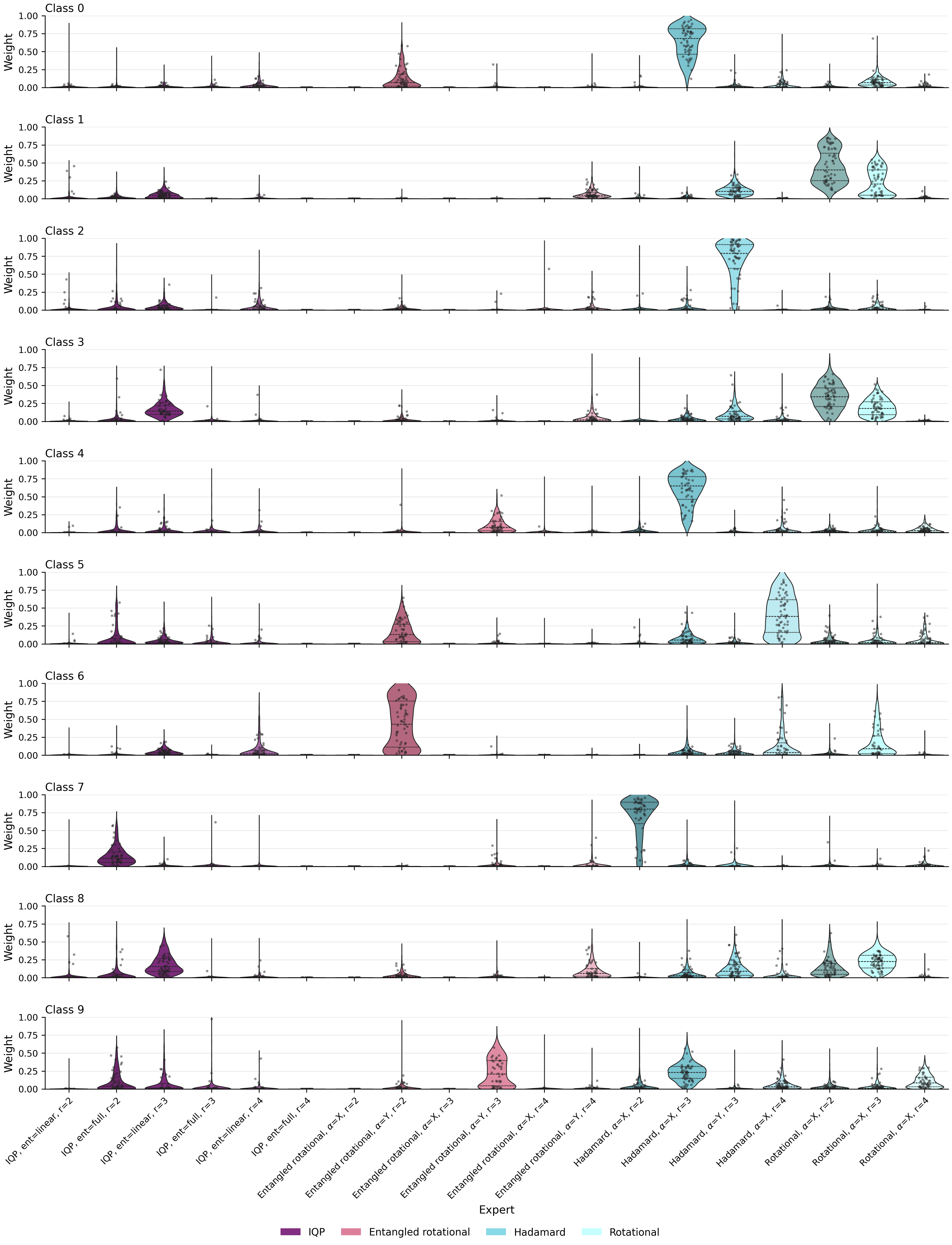}
    \caption{Distribution of the gating weights assigned to each expert over the test samples of each class, for a representative run of Soft-MoQE ($T = 0.1$, $h = 50$) on the MNIST 10-class task. Violins are coloured by expert family, each point corresponds to a test sample, and dashed lines indicate the median and the quartiles. Expert labels report the family, the entanglement or rotation axis and the number of repetitions $r$.}
    \label{fig:violin_weights}
\end{figure*}

 The performance gap with the single best expert underlines that the combination of more than one quantum embedding is beneficial over relying on any individual one. At the same time, the large gap with the uniform average shows that the classification capacity of Soft-MoQE cannot be reduced to a naive aggregation of diverse quantum representations. 
 Overall, these results show that the performance of Soft-MoQE relies on the learned weighting of the experts.
The gating network uses the expert library in a selective and class-dependent way. Figure \ref{fig:violin_weights} reports the distribution of the gating weights over the test samples of each class, for a representative run, with validation MCC closest to the median over the 30 samplings, of the best configuration ($T = 0.1, h = 50$) on the MNIST 10-class task. For most classes, the weight is concentrated on a small subset of experts, and this subset changes across classes. Hadamard experts dominate several classes, such as the expert with $X$ rotations and $r = 3$ for digits 0 and 4, the one with $r = 2$ for digit $7$ and the one with $Y$ rotations and $r = 3$ for digit $2$, while rotational experts carry most of the weight for digits 1 and 3. Entangled rotational and IQP experts are instead activated for specific classes, such as the entangled rotational expert with $Y$ rotations and $r = 2$, which dominates for digit $6$, and the IQP experts, which contribute to digits $3$, $7$, $8$ and $9$. The weights also vary among samples of the same class, indicating that the routing depends on the individual input and not only on its class. A few experts, such as the entangled rotational experts with $X$ rotations, receive negligible weight for all classes, consistent with redundancy among structurally similar encodings.  A similar class-dependent use of the experts is observed on FashionMNIST (\textbf{Figure \ref{fig:supp_weights_fm}}) where each class relies on a combination of a few experts, while on the parity task, in which each of the two classes groups several digits, the weight is distributed over a larger number of experts (\textbf{Figure \ref{fig:supp_weights_parity})}

\subsection{Baseline comparison}
\label{sec:comparison}
\begin{figure*}[t]
    \centering
    \includegraphics[width=1\textwidth]{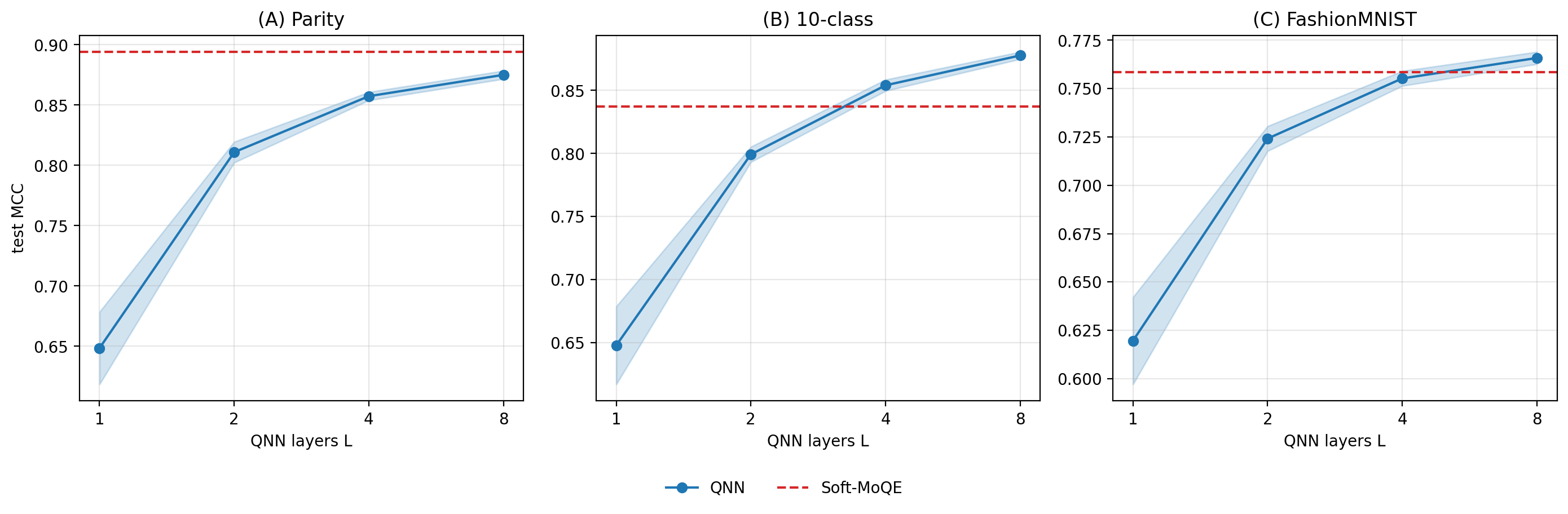}
    \caption{Test MCC of the QNN as a function of the number of layers $L$ for (A) parity, (B) 10-class digit classification and (C) FashionMNIST. Solid lines and shaded bands show the mean and standard deviation over the 30 samplings; at each $L$ the learning rate with the highest mean validation MCC is used. The dashed line indicates the mean test MCC of Soft-MoQE ($T = 0.1$, $h = 50$).}
    \label{fig:qnn}
\end{figure*}

We benchmarked Soft-MoQE against three models described in Section \ref{sec:baselines}, a linear baseline identical to the final layer of Soft-MoQE, an MLP,  and  a variational QNN in which the parameters of the quantum circuit are trained.
As reported in Table \ref{tab:performance}, the proposed Soft-MoQE model achieves higher MCC and  accuracy than the Linear baseline, confirming that the classification performance of the proposed model cannot be attributed to the final classical layer alone. The MLP, trained on the same input, provides a reference for the performance attainable by a standard classical model on the 10 principal components, and achieves the highest MCC on the 10-class and FashionMNIST tasks.
With respect to the QNN, the proposed model achieves comparable, task-dependent performances (Table \ref{tab:performance}). Soft-MoQE achieves a higher test MCC on the parity task ($0.894 \pm 0.006$ against $0.875 \pm 0.003$), lower on the 10-class task ($0.84 \pm 0.02$ against $0.878 \pm 0.003$) and slightly lower on FashionMNIST ($0.758 \pm 0.00$6 against $0.766 \pm 0.003$). These values refer to the best-performing QNN, selected on the validation set, which uses $L = 8$ layers on all tasks, corresponding to $240$ trainable quantum parameters ($P$), whereas Soft-MoQE trains no quantum parameter. We note that the performance of the QNN depends strongly on the size of the circuit. The test MCC increases with the number of layers $L$, and thus with the number of trainable quantum parameters (Figure \ref{fig:qnn}) for all three considered tasks. On the parity task even the deepest circuit considered remains below Soft-MoQE (Figure \ref{fig:qnn}(A)). On the 10-class task the QNN exceeds Soft-MoQE from $ L \ge 4$, corresponding to $P=120$, while on FashionMNIST it reaches a comparable MCC at $L = 4 $ and exceeds it only at $L = 8$, with $P= 240$  (Figure \ref{fig:qnn}(B) and (C)).  
\begin{table*}[t]
\centering
\caption{MCC and accuracy on the MNIST parity, MNIST 10-class and FashionMNIST test sets for Soft-MoQE ($T = 0.1$, $h = 50$), the variational QNN ($L = 8$), the linear baseline and the MLP. All models received the same 10 principal components as input and were trained on the same 30 random training/validation samplings. Results are reported as mean $\pm$ standard deviation over the samplings.}
\resizebox{0.95\textwidth}{!}{
\begin{tabular}{@{}lcccccc@{}}
\toprule
\multicolumn{1}{c}{\multirow{2}{*}{\textbf{Model}}} &
\multicolumn{2}{c}{\textbf{MNIST: Parity}} &
\multicolumn{2}{c}{\textbf{MNIST: 10-class}} &
\multicolumn{2}{c}{\textbf{FashionMNIST}} \\
\cmidrule(l){2-7}
 & MCC & Accuracy & MCC & Accuracy & MCC & Accuracy \\
\midrule
Soft-MoQE & $0.894 \pm 0.006$ & $0.947 \pm 0.003$ & $0.84 \pm 0.02$ & $0.85 \pm 0.02$ & $0.758 \pm 0.006$ & $0.782 \pm 0.005$ \\
QNN & $0.875 \pm 0.003$ & $0.938 \pm 0.002$ & $0.878 \pm 0.003$ & $0.890 \pm 0.003$ & $0.766 \pm 0.003$ & $0.789 \pm 0.003$ \\
Linear baseline & $0.640 \pm 0.001$ & $0.8196 \pm 0.0007$ & $0.7806 \pm 0.0008$ & $0.8026 \pm 0.0007$ & $0.715 \pm 0.001$ & $0.743 \pm 0.001$ \\
MLP & $0.867 \pm 0.007$ & $0.934 \pm 0.004$ & $0.893 \pm 0.004$ & $0.903 \pm 0.003$ & $0.778 \pm 0.004$ & $0.802 \pm 0.003$ \\
\bottomrule
\end{tabular}
}
\label{tab:performance}
\end{table*}
However, more trainable parameters come with a higher computational cost in terms of quantum resources, while Soft-MoQE relies on fixed circuits, which significantly reduces the number of circuit executions required for training. Specifically for a library of $n$ experts, Soft-MoQE  requires $n \times N$ circuit executions for a dataset of $N$ samples. These embeddings are shared by all training samplings and hyperparameter configurations.

 The QNN instead requires circuit evaluations at every optimization step. On quantum hardware, estimating the gradient with the parameter-shift rule requires $2P + 1$ executions per sample~\cite{mitarai2018,schuld2019}, so that a single training run costs about $(2P + 1) \times N_{\text{train}} \times E$ executions, with $N_{\text{train}} $ training samples and $E$ training epochs. For the experiments reported here, with $n = 19$ experts, $N = 70000$ samples, $N_{\text{train}}  = 50000$ training samples and $P = 240 $ for the best QNN configuration, Soft-MoQE requires about $1.3 \times10^6$ circuit executions in total. On the other hand a single training epoch of the QNN already requires about $2.4 \times 10^7 $executions. Since in our experiments the QNN is trained for $156$ to $243$ epochs on average, depending on the task, a full training run amounts to between $3.8 \times 10^9$ and $ 5.8 \times 10^9$  executions, without counting the hyperparameter search. Since all models were simulated, these numbers are analytic estimates of the cost on quantum hardware, obtained assuming the same number of shots per execution for both models. On the other hand at inference Soft-MoQE requires $n$ circuit executions for each new sample whereas the QNN needs only one, an overhead that nevertheless remains small compared with the cost of training.

\section{Conclusions}
In this work we introduced Soft-MoQE, a hybrid architecture in which the quantum component consists of a library of fixed feature maps and learning takes place entirely in a classical gating network that routes each input across them. The best performance was obtained with a moderately selective routing ($T = 0.1$), and comparison with the single best expert and with the uniform average of all experts shows that the classification performance of Soft-MoQE emerges from the learned weighting of the experts, and not from any individual expert or from their uniform combination. Furthermore, the selection patterns depend on the class label, with different subsets of experts used for different classes.

Soft-MoQE also outperformed a linear classifier identical to its output layer, confirming that its performance does not reduce to the final classical layer. Compared with a variational QNN, it reached comparable, task-dependent performance without training any quantum parameter, and with a quantum cost of training lower by orders of magnitude, since each fixed circuit is executed only once per sample.

While our results show that a learned routing over fixed quantum feature maps is effective, we acknowledge that the library considered here, built from a few standard families of encodings, is not exhaustive, and that the most useful experts are likely to depend on the task.
Furthermore, the experts used in this work are structured feature maps without internal parameters. 
Including fixed quantum circuits with randomly drawn parameters as experts, building on quantum extreme learning machines \cite{mujal2021opportunities}, would extend the library beyond hand-designed encodings, toward more entangled and potentially harder-to-simulate representations, while preserving the fully classical training of Soft-MoQE. This is a direction we plan to explore in future work.
Finally, the numerical experiments are restricted to noiseless simulations on 10 qubits; since the architecture does not depend on the size of the experts, its extension to larger circuits requires no change to the model, while its performance at larger qubit numbers and its behaviour on noisy quantum hardware remain to be assessed.
Overall, building on the idea that combining multiple quantum models can be particularly effective in the QML setting \cite{schuld_quantum_2018,incudini2023resource}, Soft-MoQE introduces a class of modular architectures where fixed quantum feature maps are combined into an adaptive learning framework that integrates the classical component in a controlled and interpretable way, mitigating the trainability challenges of PQCs. In this perspective, the central operation is no longer the variational optimization of quantum parameters, but the learned routing among a set of candidate quantum circuits.

\section*{Acknowledgement}
We acknowledge ISCRA for awarding this project access to the LEONARDO supercomputer, owned by the EuroHPC Joint Undertaking, hosted by CINECA (Italy). We also acknowledge the ICSC for awarding this project access to the EuroHPC supercomputer LEONARDO, hosted by CINECA (Italy).
\newline
\section*{Declarations}
\subsection*{Funding}
The European Union - Next Generation EU, in the context of PNRR, Investment 1.5 Ecosystems of Innovation, Project Tuscany Health Ecosystem (THE), ECS00000017, Spoke 3. CUP: B83C22003930001 (to VR and RDA). Additional financial support was obtained from the Italian Ministry for Education, University and Research under the programme “Ricerca Locale ex-60\%” and PNRR M4C2 HPC—1.4 “CENTRI NAZIONALI”—Spoke 8
\subsection*{Conflict of interest}
The authors have no competing interests to declare that are relevant to the content of this article.
\subsection*{Data and code availability }
The data, code, and experimental results associated with this project are available on GitHub at the following link: \url{https://github.com/repettovaleria/Soft-MoQE}
\subsection*{Authors contributions}
VR and PF conceived the study. VR performed the computations and analyses, interpreted the results and drafted the manuscript. RDA acquired the funding, provided the computational resources, supervised the work, interpreted the results and reviewed the manuscript. PF supervised the study, interpreted the results and reviewed the manuscript. All authors approved the final manuscript.

\printbibliography

\clearpage
\appendix
\setcounter{figure}{0}
\setcounter{table}{0}
\renewcommand{\thefigure}{S\arabic{figure}}
\renewcommand{\thetable}{S\arabic{table}}
\renewcommand{\theHfigure}{S\arabic{figure}}
\renewcommand{\theHtable}{S\arabic{table}}
\section{Gating weights on MNIST parity and FashionMNIST}
\label{sec:supp_weights}

\begin{figure}[h!]
    \centering
    \includegraphics[width=\textwidth]{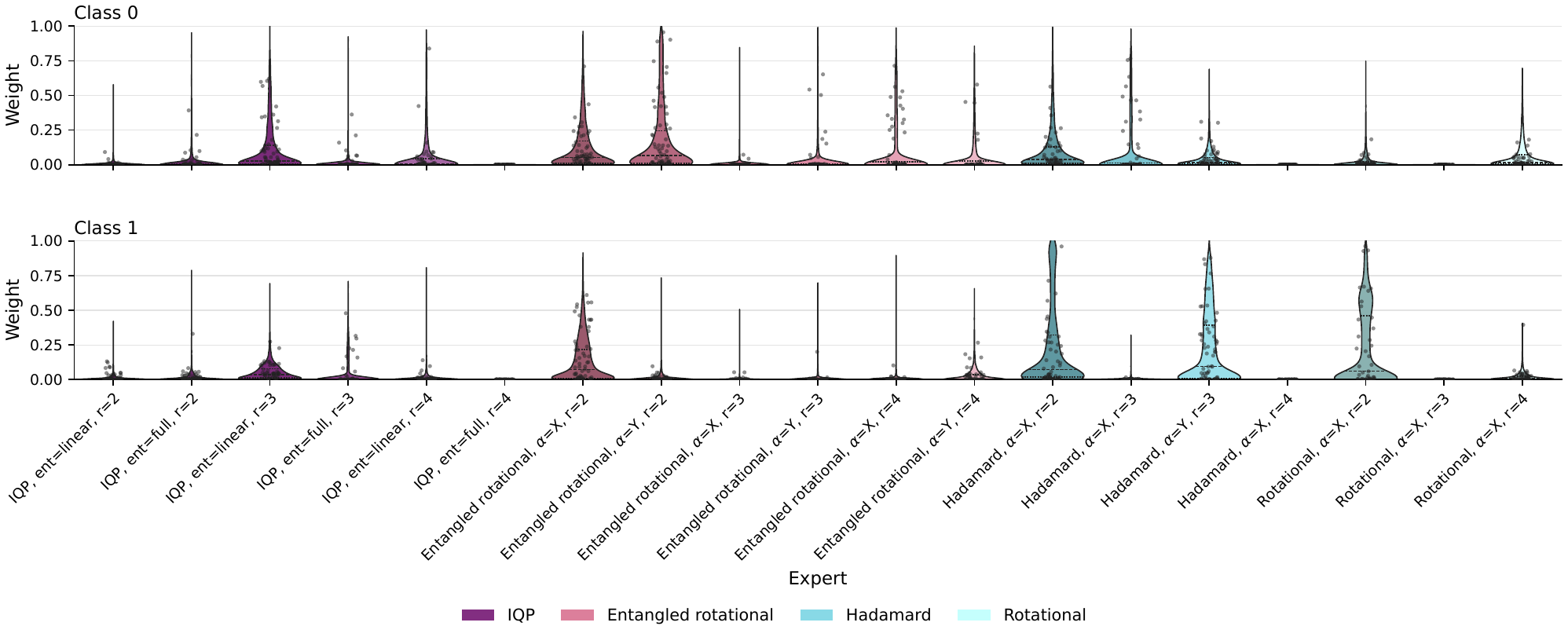}
    \caption{Distribution of the gating weights assigned to each expert over the test samples of each class, for the run of Soft-MoQE ($T = 0.1$, $h = 50$) with validation MCC closest to the median over the 30 samplings, on the MNIST parity task (class 0, even digits; class 1, odd digits).Violins are coloured by expert family, each point corresponds to a test sample, and dashed lines indicate the median and the quartiles. Expert labels report the family, the entanglement or rotation axis and the number of repetitions $r$.}
    \label{fig:supp_weights_parity}
\end{figure}

\begin{figure}[p]
    \centering
    \includegraphics[width=\textwidth,height=0.85\textheight,keepaspectratio]{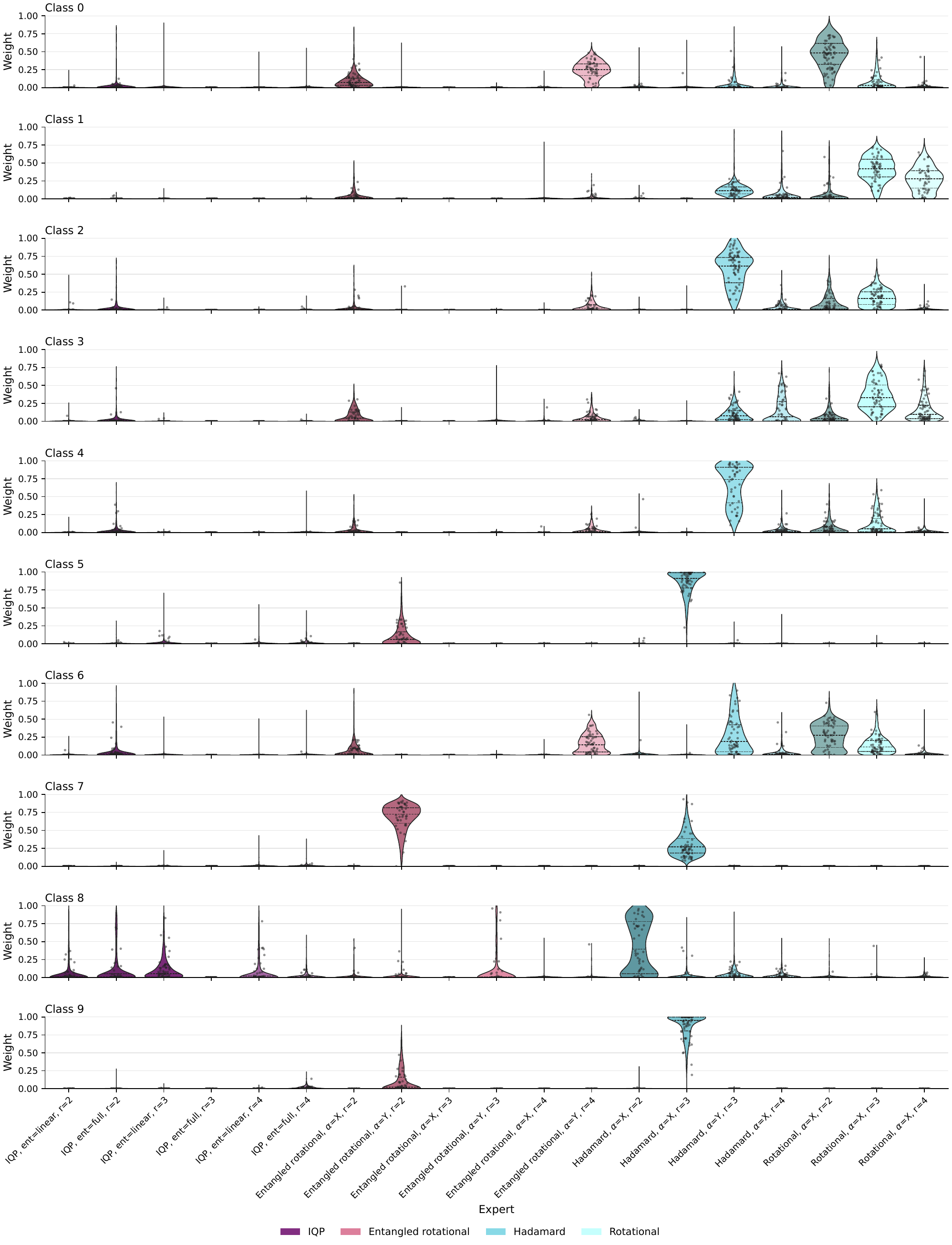}
    \caption{Distribution of the gating weights assigned to each expert over the test samples of each class, for the run of Soft-MoQE ($T = 0.1$, $h = 50$) with validation MCC closest to the median over the 30 samplings, on FashionMNIST. Violins are coloured by expert family, each point corresponds to a test sample, and dashed lines indicate the median and the quartiles. Expert labels report the family, the entanglement or rotation axis and the number of repetitions $r$.}
    \label{fig:supp_weights_fm}
\end{figure}

This section reports the distribution of the gating weights for the MNIST parity and FashionMNIST tasks, complementing the analysis of the MNIST 10-class task in the main text. Consistently, we consider the run of the best configuration ($T = 0.1$, $h = 50$) whose validation MCC is closest to the median over the 30 samplings. On the parity task (Figure~\ref{fig:supp_weights_parity}), in which each of the two classes groups several digits, the weight is distributed over a larger number of experts. On FashionMNIST (Figure~\ref{fig:supp_weights_fm}) each class relies on a combination of a few experts, and this combination changes across classes.

\end{document}